# A Three Steps Methodological Approach to Legal Governance Validation

Pompeu CASANOVAS [a,b,c], Mustafa HASHMI [a,c], Louis DE KOKER [a,d],
Ho-Pun LAM [e]

[a] *La Trobe Law School, La Trobe University, Melbourne, Australia*
{M.Hashmi, L.DeKoker}@latrobe.edu.au
[b] *Artificial Intelligence Research Institute of the Spanish National Research Council (IIIA-CSIC) Barcelona, Spain*
pompeu.casanovas@iiia.csic.es
[c] *UAB Institute of Law and Technology, Autonomous University of Barcelona*
[d] *University of the Western Cape, Bellville, South Africa*
[e] *Department of Computing, Xi'an Jiaotong-Liverpool University, Suzhou Industrial Park, Suzhou, 215123, Jiangsu, China.*
ho.lam@xjtlu.edu.cn

To Aaron Cicourel (1928-2023), *in memoriam*

**Abstract.** We present in this position paper a methodology to validate legal governance regulatory models from an empirical approach, as illustrated by means of three diagrams: (i) a scheme drawing the rule and meta-rule of law; (ii) a metamodel for legal governance; (iii) a causal validation scheme for legal compliance. These visualisations refer to different sets of notions corresponding respectively to (i) a general scheme with three dimensions and four clusters, (ii) a meta-model encompassing *legal compliance through design* (LCtD) and *ecological validity*, and (iii) the construction of an empirical validation model of causal chains. The final aim of the methodology is to build and test *smart legal ecosystems* (SLE) for Industry 4.0 and 5.0.

## 1. Introduction

This position paper outlines an approach to validate legal governance models, i.e., to validate the results of conditions and the interrelationship among them, and to generate legal ecosystems in the so-called Internet of Things (IoT), the Web of (Linked) Open Data (LOD), and Industry 4.0 (I4.0) and 5.0 (I5.0). This approach is meant to bridge the different technologies involved. I4.0 refers to smart manufacturing, covering a wide range of production and distribution processes. I5.0 refers to the human social effects and consequences of adopting smart manufacturing on the Internet of Things, including ethical and legal values and compliance, thus, linking automation and the effective use of cyber-physical systems to the human dimension [1] [2].

It should be noted that this approach can also be located in between regulatory implementation models focusing on private and/or on public law. It can be easily expanded to other sectors in which distributed (or federated) architectures and monitored semi-automated asymmetric multi-level governance are needed, such as cybersecurity, health, financing, and banking. Our approach embraces a conceptual and metricised gradual perspective, considering different types of compliance—such as strict, partial, over, and non-compliance [3] [4] [5].

This is not the first time that we consider solutions in this field. Three of the authors have been partially applying this approach and addressing requirements and issues on legal theory, ethics, and regulatory models in several cybersecurity[1], immigration[2] and now, industry projects[3]. Two of the authors have been intensively working on compliance rules modelling techniques, defeasible logic reasoners such as SPINdle[4], and compliance checking systems such as REGOROUS[5]. We are also researching a comprehensive *empirical* methodology to analyse legal documents, behaviour, practices, instruments, and sources simultaneously in an integrated manner. In this sense, this is a research agenda, focused on broad problems and methodologies and approaches to work towards solutions.

Thus, this work is aimed at raising, if not yet solving, legal governance challenges in the emerging context of *hybrid* [6] or *symbiotic intelligence* [7], where humans and artificial agents cooperate to produce emergent second order phenomena 'that involve groups of agents who reason and decide, specifically, about actions – theirs or others' – that may affect the social environment where they interact with other agents' [7].

This is a conceptual paper, presenting the main blocks of the methodology. At this stage, we are not introducing metrics nor coding to embed them into platforms or cyber-physical systems. These tasks correspond to an implementation stage in which we can build solutions using (i) deontic logic, (ii) and thresholds to estimate the degree of compliance. We are now focused on outlining the validation three-step process model that is prior to any formalisation. Moreover, defining what can and cannot be formalised is not trivial. This also corresponds to an implementation stage in which the evaluation is carried out considering the information flows already generated through the modules built upon the different technologies connected and integrated into the architecture design (e.g. middleware and blockchain solutions).

---

[1] Cf. CAPER: *Collaborative information, Acquisition, Processing, Exploitation and Reporting for the prevention of organised crime*, https://cordis.europa.eu/project/id/261712; SPIRIT: *Scalable Privacy preserving Intelligence Analysis for Resolving Identities,* https://cordis.europa.eu/project/id/786993,; and the projects on cybersecurity held by the Australian Government Program D2D CRC: *Data to Decisions Cooperative Research Centre,*
 https://www.latrobe.edu.au/cdac/research/research-projects/data-to-decisions-crc    Cf. especially, *DC25008: Compliance by Design (CbD) and Compliance through Design (CtD) solutions to support automated information sharing (2018-19). Law and Policy. Project C. Spent Convictions Use Case.* Australian Government funded Data to Decisions Cooperative Research Centre (2018-2019), end-user: Australian Criminal Intelligence Commission. https://zenodo.org/records/3271525

[2] ITFLOWS: *IT tools and methods for managing migration FLOWS,* https://cordis.europa.eu/project/id/882986 .

[3] OPTIMAI: *Optimizing Manufacturing Processes through Artificial Intelligence and Virtualization*  https://optimai.eu/

[4] http://spindle.data61.csiro.au

[5] https://research.csiro.au/data61/regorous/

The remainder of the paper is as follows. Section 2 introduces the previous work on this subject, some definitions and three remarks. Section 3 is divided into three subsections putting in place the three steps methodology for legal governance and smart legal ecosystems evaluation. Section 4 draws some conclusions and describes the future work.

**2. Compliance and the Internet 5.0**

*2.1. Some preliminary definitions*

By *legal governance* we understand the set of processes that generate a sustainable regulatory ecosystem reflecting fundamental legal concepts of a modern democracy [8]. We conceive it as an explanatory and validation notion, primarily informed by a social and cognitive science approach, to support the implementation of the rule of law in hybrid environments in which Human/Machine/Interaction (HMI) and Human/Robotic/Interaction (H/R/I) constitute symbiotic contexts and scenarios.

A *legal ecosystem* can be defined as a complex and dynamic system that includes multiple levels of governance, ranging from local to national and international, and involving a wide range of actors, including lawmakers, judges, lawyers, law enforcement officials, civil society organizations, companies, corporations, and ordinary consumers and citizens [3] [9].

A *smart legal ecosystem* works in an intelligent environment, encompassing the features of the IoT 4.0 and 5.0 (ethics and law) bringing about legal compliance on real time, and being (partially) embedded into cyber-physical systems [10].

*Compliance*, in a broad sense, can be understood as fulfilling or aligning with regulatory constraints [9]. Regulatory compliance points at a previously selected set of requirements for industry and business and industry processes, as set e.g., by ISO/IEC 27002, among many others. By *legal compliance* we broadly mean the whole process of fulfilling the requirements contained both in traditional legal instruments (mainly hard law, i.e. the outcomes of Parliaments and Courts), and other regulatory instruments (such as soft law standards, best practices, and policies, etc.). These instruments are best designed following the processes proposed in the EU *better regulations guidelines* [11] *and toolbox* [12]. Broader legal compliance, to be distinguished from technical legal compliance as an IO4 and IO5 toolkit, also entails a mindset and social behaviour is a component and a result of this broader compliance process.

*2.2. The emergence of new legal instruments*

New regulatory and legal instruments are complementing those that we inherited from a non-distant past. But, as Bob Johansen [13] puts it, we are facing a brave new world, a VUCA world: *Volatile, Uncertain, Complex, and Ambiguous*. Interestingly, Johansen borrowed this term from the Army War College in Carlisle, Pennsylvania. It was used by young officers, and he expanded it to market and business innovation and leadership processes: 'We are on a twisting path toward—but never quite reaching—a place where everything will be distributed. This path will be characterized by increasing speed, frequency, scope, and scale of disruption.' [13, p. vii]. Alternative but complementary visions following the human rights line of Cathy O'Neil [14] are stressing the social and political dark side of this process. '*AI systems are ultimately designed to serve existing dominant interests. In this sense, artificial intelligence is a registry of power.*' [15]

This is why it is so important not to throw the baby out with the bathwater. In the VUCA world, the rights, and principles of the rule of law should be preserved and enhanced, mainly using the same techniques and technologies that could be operated to diminish them. Likewise, controls should be put in citizens' hands (not relying only on state oversight).

Before and during the pandemic, the quest for new legal instruments to build up dynamic systems started anew, especially in medicine, health, and global and international law. Online certification procedures, counselling, negotiation, dialogue, medical attention, and Online Dispute Resolution tools (ODR) took off with renewed energy, as noticed by *The Lancet* [16]. It situated compliance, again, at the centre of the legal implementation process conceiving it in relation to a classic understanding of what international law consists of, i.e. relationships between states. This is close to our concept of *legal governance*, although, in our view, this notion includes the regulatory behaviour of all relevant stakeholders, not only the regulatory activities carried out at the state and inter-state levels (parliament, government, administration, judiciary).

From our point of view, the double implosion experienced by the legal profession in several stages of the globalisation process and, lately, the emergence of legal web services should be considered to explain the developments of these new legal instruments [17] [18]. Other approaches face the new legal instruments in the light of meta-regulation, i.e. 'the rules that govern how individual policies are developed and reviewed', e.g., among other, impact assessment, stakeholder consultation, and evaluation [19]. This latter multi-levelled perspective entails validation processes, which is one of the main topics of our approach.

### 2.3. Epistemic foundations

The methodology that we are proposing enhances *situated* cognition, technology, and regulations, following some advances in cognitive and social sciences, neuroscience, and deontic philosophy partially based on the pioneering work carried out by Edmund Husserl and his influence on Karl Bühler, Eric Voegelin, and Alfred Schütz [20]. These interwar developments constitute a specific trend within the phenomenological tradition, linking hyletic (sensitive) knowledge with the emergence of environments in specific (pragmatic) contexts. Expression is 'a parable of action', according to Engel and Bühler [21]. They could develop and discuss it on personal bases, without excluding Hans Kelsen and other neo-Kantian normative theorists from this discussion [22]. In the next generation, social and computer scientists drew on these cognitive notions—relationships, interactions, environments— in a range of fields including computer science [23], linguistics [24], anthropology [25], or sociology [26].[6]

---

[6] According to Cicourel, e.g., 'The general point is that the communication we attribute to discourse and any paralinguistic and nonverbal activities is part of a complex, multi-level, not always integrated setting. Multiple sources of information are always operative and so our analysis of

The concept of *user-centered system design* was introduced by Don Norman and Stephen Draper in 1986 [27]. Evidence for the emergence of collective and *distributed cognition* was empirically furnished by Edwin Hutchins in 1995 [28], and expanded by Holland, Hutchins and Kirsch in 2000 [29]: 'Unlike traditional theories, [the theory of distributed cognition] extends the reach of what is considered *cognitive* beyond the individual to encompass interactions between people and with resources and materials in the environment.' [ibid. pg. 175]. These are the foundations of what now is called the *human-centered design of artificial intelligence* [30].

The exogenous variables depend on the selected level of abstraction for building and applying the regulatory design. Thus, we can combine this cognitive distributed approach with the inferential rule modelling which can also be implemented as a component of the regulatory model. This combination does not prevent us from looking norms from the outside as well. On the contrary, humans (and robots) do not solely interpret the content of norms. They *play* with them, they figure out how they look like, they create and recreate their types and instantiations in many ways.

Agents, be they human or artificial, are situated in an interactive dynamic *nomotropic* space in which norms and rules can be understood from a behavioural point of view, and this behaviour can divert from just complying or violating the rules [31]. It can recreate, reformulate, or even rewrite them as entities, as language, or as mere objects. *Acting-in-function-of rules* is not acting according to their content but considering the possibility of reshaping, reusing, or ignoring them in accordance with a plurality of interests, including a *contrario* interpretations of their explicit meaning. It is worth noting that we can find a similar perspective in the early developments of Multi-Agent Systems and artificial societies. Agents can also cheat and lie. Autonomous goal-directed behaviour as the root of all social phenomena has been one of the guiding main objectives of Cristiano Castelfranchi's contributions at micro and macro levels [32].

### 2.4. Previous work: A legal quadrant for the rule of law

Adopting these epistemic cognitive grounds helped us to better formulate the notion of substantive rule of law in such a way that could be represented and applied through formal languages. Between 2017 and 2021 we developed (i) a regulatory quadrant to represent the rule of law; (ii) a cluster of concepts to describe instruments and processes of the law; (iii) the methodology followed to select technical papers concerning regulatory compliance; and (iv) an initial mapping to frame the selected papers about legal compliance that we used in a final survey (on nearly 900 articles). The result was plotted against a conceptual clustering that we found useful for analysing and differentiating between *Compliance by Design* (CbD) and *Compliance through Design* (CtD). We concluded that CbD and CtD should be treated separately, as legal compliance and business compliance do not always refer to the same concepts and requirements. We summarised our previous results in [33]. Figure 1 reproduces the legal quadrant that we drew and used as a compass for several research projects.[7] It shows how the validity of norms (i.e.

---

discourse must necessarily simplify or reify many aspects of social interaction as well as what we are calling discourse.' [26, p.101]

[7] We coded the literature and derived a codification protocol to meet the objectives of the analysis. In the coding process, we used a sample of the most frequently used concepts—we created 327 nodes across four clusters of distinct themes according to the quadrant hard law, ethics, policies, and soft law. Along these lines, we also created 157 additional relationship nodes, expanding the

their 'legality') emerges from four different types of regulatory frames, with some distinctive properties. Properties are understood here as correlating dynamic patterns. We identified four basic components for the societal implementation of the rule of law — *hard law, soft law, policies, and ethics*— and the relationship between them. We considered the sources, domains, and relationships with respect to citizens (interconnectedness of norms or rules).

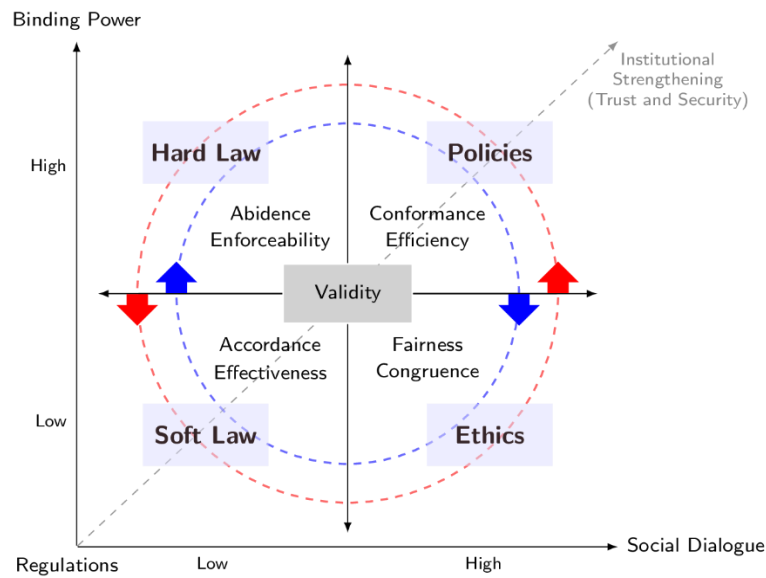

**Figure 1**: Legal quadrant for the rule of law. Source: [33].

**3. Three Steps Methodological Approach for Legal Governance Validation**

Now we will sketch our methodology, identifying its main components and presenting them in ordered sequences. Nevertheless, a full development and theoretical discussion will be not yet offered. Our intention is to provide a short first summary, and to find the main research questions that should be answered in the next future.

*3.1. First Step: A Meta-rule of Law Scheme*

Figure 2 provides a general schematic representation of the rule of law and its counterpart, the *meta-rule of law*, i.e. the embedded protections of the substantive rule of law

---

analysis to 484 nodes. The coding process resulted in a matrix of nodes reflecting the interactions of various concepts and dependencies between them. We applied (i) the Pearson's coefficient correlation, (ii) Jaccard's statistical techniques to investigate the relationships between the concepts, and across inter/intra clustered themes, (iii) and we also used the Sørensen similarity coefficient to compare them and validate the similarity and strength of the relationship between the concepts. The interested reader is kindly requested to go to [33] to find the details, further references, and open discussion.

in computer systems through formal languages. It highlights the difference between regulations that were conceived to rule human social behaviour, and the new digital dimension in which rules, principles and instruments are embedded into formal languages and computational codes to be digitally generated, interpreted, and implemented. Natural, semiformal, and formal languages have different properties. As shown by the *ergativity* of polysynthetic (not Indo-European) languages, there is no universal grammar covering all aspects of expressive verbal morphology [34].

The cycle of the meta-rule of law is plotted on Figure 1. It shows two axes (vertical: binding power, horizontal: social dialogue), three dimensions (social, legal, and computational), four clusters (hard law, policies, soft law, and ethics), and four cornerstones (multi-stakeholder governance, anchoring institutions, the binomial trust/security, and institutional strengthening) to produce regulatory effects. All these elements are components of the regulatory system lifecycle, i.e. elements of *legal governance*. We considered the implementation of the rule of law along two related dimensions at the empirical level: (i) (binding) institutional power—the vertical axis in the quadrant— and (ii) social dialogue (negotiation, compromise, mediation, agreement)—the horizontal axis in the quadrant. The semi-automation of legal governance is the next step, i.e. the creation of a regulatory interspace, bringing together all relevant stakeholders (including rulers, industry, and citizens), and the AI and legal instruments at their disposal.

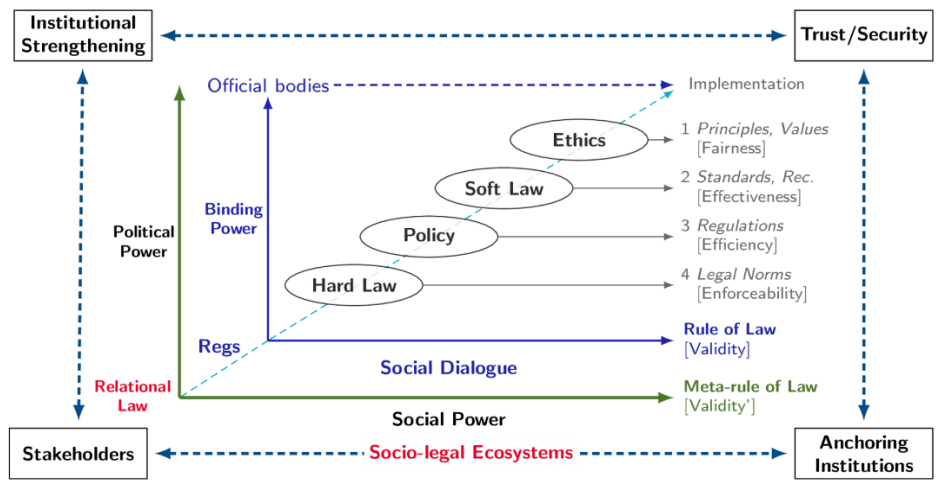

**Figure 2.** Scheme for the meta-rule of law. Source: [3] [8].

From the empirical approach that we are adopting here it should be noted that 'validity' (as a synonym of 'legality') is a second-order property that emerges only when a threshold for enforceability, efficiency, effectiveness, and fairness have been stablished and applied. It is not applied to norms, but to the whole regulatory model as a system.[8] When plotted on a computer language, we can predicate validity as a formal property related to consistency (*validity'*), but it does not drag 'validity' with it in the same way that 'truth' does in descriptive logic reasoning. Our contention is that to make it 'legal', at the empirical level (i.e., at the *existential* level, not at the deontic one) more requirements are needed related to a more complex compliance process; and as we will state in the next sections, *validation* processes cannot be equated with legal *validity* (as a second-order property or as a means of achieving consistency on the regulatory model).

---

[8] Conte [35] criticizes the unicity of the notion and contends that the term 'norm' refers at the same time to at least five different things: a deontic enonciation, a deontic proposition, a deontic statement, a deontic state of affairs, a deontic *noema*. We can observe that these 'deontic entities' are working in contexts that are deemed to be also different.

Likewise, extracting (formal) rules from norms formulated in natural language is a problem that has not completely been solved either. In the early times of AI and Law it was known as the *legal isomorphism* problem [36] and, even now, extracting normative information from legal documents is still a challenge. Hence, after listing several current methodologies based on Natural Language Processing (NLP) or Machine Learning (ML) techniques, Hashmi et al. [37] contended:

> "[…] in our view, the norms extraction process is far deeper than just extracting the document structure and classifying the terms but identify and extract deontic components of rules, and correctly assign the terms to the antecedent and the consequent of the rules. Also, extract the co-reference links that are present in the legal documents, align the terms that are used in the legal text and the terms that we want to use in the rule providing thus, a unified representation of the norms for further formalisation. We strongly believe that the proper extraction of norms is an ongoing challenge and does not seem to be fully automated in near future. However, we also believe that due to the complexity of the legal texts and time required to manually extract norms, (even partially) automating this task would be beneficial."

In addition, we should add the difficulty of grasping and defining the *emergence* of rules straight from the interactive behaviour of agents instead of documents or written provisions. There are many different possibilities to build and describe them from a collective point of view. Ostrom [38] defined several ways of describing *shared strategies*, and so did Ghorbani et al. [39] for Multi-Agent's Systems (MAS) behaviour. We should differentiate several problems: (i) rule extraction (from norms); (ii) norm extraction (from documents); (iii) rule and norm extraction from interactive behaviour (shared strategies); (iv) *and* pre-modelling or conceptual model building out of shared or accepted sources (not all written). The scheme presented here, jointly with the legal quadrant, is a simple way of clustering not just legal provisions but relevant social behaviour when building a regulatory legal model.

*3.2. Second Step: A Metamodel for Legal Governance*

The second step comprises *Legal Compliance through Design* (LCtD) [9]. LCtD encompasses legal interpretation and decision-making, bridging the path from the four clusters previously identified (the sources of law) to legal governance. There are three blocks to be considered. The first one stems from the selection of sources and legal material described in the first step. The second one is focused on validity and LCtD. The third one generates the ecological validity of the regulatory model. The meta-model drawn in Figure 3 plots the whole process.

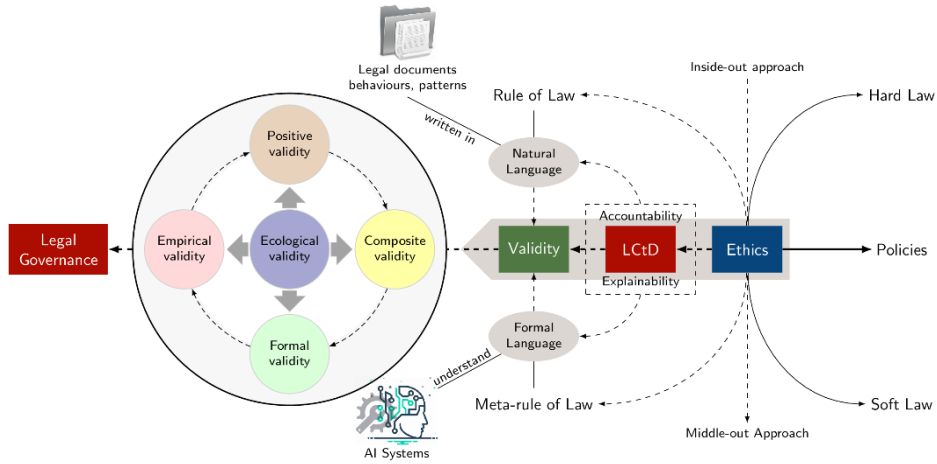

**Fig. 3.** Metamodel of Legal Governance

### 3.2.1. Ethics, LCtD and ecological validity

The first block situates Ethics at a filtering intermediary position because it also applies to AI devices, platforms, modules, and applications, independently of jurisdictional and sovereignty principles and restrictions. As a matter of fact, there is a myriad of ethical principles tailored for AI.[9] Among many other proposals, AI4People [40] suggested in 2018 the following ones: (1) *beneficence*, as promoting well-being, preserving dignity and sustaining the planet; (2) *non-maleficence*, as privacy, security and capability caution; (3) *autonomy*, as the power to decide; (4) *justice*, as promoting prosperity and preserving solidarity; and (5) *explicability*, by enabling the other principles through intelligibility and accountability. In 2019, Floridi et al. [41] differentiated explicability from *explainability*. And the same year, AI4People completed its work with some more principles and a toolkit for AI governance from the legal point of view [42]. A *middle-out approach* was proposed, to avoid the reduction of regulations to a bottom-up or top-down implementation [43]. Again, this was complemented a bit later with the *inside-out approach*, to make sure that regulations and legal instruments could be designed in a modular and scalable way as platform regulatory drivers [3].

It is worth mentioning that socio-technical systems, the coordination of Multi-Agent Systems, and Cyber-physical Systems rely on continuous informational flows at three different layers—the perception, network, and application layers. From a theoretical point of view, this third technological dimension adds more complexity to the notions of *normative and empirical validity* that have been separated into two separate fields by many legal and socio-legal theorists (from Max Weber to Robert Alexy). In contrast, we are focusing onto the validation process in an empirical chain, requiring approaches that are not reflected in the current leading theories of socio-legal or legal validity.

---

[9] Cf., e.g. the Asilomar Principles for AI, https://futureoflife.org/open-letter/ai-principles/ and the IEEE Principles, https://standards.ieee.org/wp-content/uploads/import/documents/other/ead_general_principles.pdf

LCtD leads to the emergence of *ecological validity* (a tuple of positive, empirical, composite, and formal validity) [8] [9]. Positive validity refers to the social acceptance of a shared regulatory framework. Empirical validity refers to the degree of implementation of the model. Composite validity is the compliance statistical indicator (index) that can be built from the degree of effectivity, effectiveness, fairness, and empirical validity measured from a defined threshold or estimator. Formal validity denotes the internal formal consistency of the model. Only from their combination can emerge the *ecological validity* that is necessary to trigger legal governance (i) to embed protections into the system; (ii) to empower stakeholders (citizens, consumers, organisations, communities, etc.), (iii) to protect and enhance their individual and collective rights, i.e. *to create the conditions for generating a sustainable legal ecosystem.*

*3.2.2. Example: Smart Manufacturing*

The elements of the three dimensions relevant in this context (social, legal, and technological) matter from a technical perspective, as they must be computed in real time or near-real time. *Validation* occurs in the technological dimension, between the social and the legal dimensions, as a separate process but uniting and linking the two former dimensions. As we will see, the meta-model of legal governance envisages legal compliance validation processes that occur in real time and in parallel. Thus, the legal ecological validity is generated by means of a CtD process at the time that a first technological validation is also produced. This can be possible because what is effectively generated is a hybrid HMI sustainable legal ecosystem, and not solely a system of norms holding abstract properties. But this is the challenge: *How and at what stage can interpretive decisions be combined with real time execution of compliance processes?* The question must be addressed any time a model is built to regulate a specific ecosystem generated by an information system and their human counterparts (be they end users, controllers, or managers).

The 'smart factory' may be a helpful example. A 'smart factory' refers to the vertical integration of various components to implement a flexible and reconfigurable manufacturing system [44] [45]. OPTIMAI is an I4.0 EU project to create a Decision Support Framework for the EU industry. The OPTIMAI framework consists of a self-organized multi-agent system assisted with big data-based feedback and coordination. As described by its designers, the model includes an intelligent negotiation mechanism for agents to cooperate with each other. Its architecture has been introduced in a functional way as:

> the OPTIMAI project architecture for zero-defect manufacturing (ZDM), applicable to a variety of industrial verticals. To *realise a standards-based approach*, we elaborate on the parallels drawn between the presented architectural framework and two leading reference architectures underpinning the "factories of the future" vision (RAMI 4.0 and IIRA). System specifications for ZDM are hence defined according to the perspectives of the two architectural models, allowing us to examine cutting-edge technologies for ZDM (such as blockchain, AI and AR) as both an I4.0 solution, as well as an Industrial Internet of Things system. [44]

Standards are applied through architecture and modular building, embedding them as functional requirements of the entire system, and keeping humans in the loop. From a control engineer's perspective, the smart factory, it has been said, can be viewed as a dual closed-loop system: 'One loop consists of physical resources and cloud, while the second loop consists of supervisory control terminals and cloud' [45].

The validity and traceability of transactions are ensured by: (i) the decentralization produced by an authorized blockchain with an access control layer; (ii) the use of Ethereum with the Proof of Authority (PoA) consensus mechanisms; (iii) the smart contracts executed between the participants; (iv) the middleware that controls access and provides the data to the blockchain [46]. Hence, a smart regulatory ecosystem can be produced through the dataflows, *operating in real time*.

Likewise, a Smart Legal Ecosystem (SLE) can also be generated, but not in a direct way: it requires a further validation process to ensure that the transactions brought about by the system *are not only agreed and valid, but legal*. SLE emerges (rather than supervenes) from the collective coordination of HRI interactions, and this is what should be checked out and tested, i.e. evaluated, as well. As already noticed in the AI & Law literature, the problem is that despite using the term *contract*, authorized blockchain and smart contracts are technological devices that cannot be deemed 'legal' *per se* [47] [48]. They are not identical to *legal agreements* [49].

Thus, our point is that there is a *third normative loop*, accompanying the online processing and generating the smart legal ecosystem that assumes a nested ecological validity of its regulatory components. The metamodel of Figure 3 must be anchored into specific regulatory models, starting with the selection of the legal instruments plotted on Figure 1. To make it happen we can use existing generative AI tools and LLMs [50] as long as we proceed in a controlled manner. In the same way, we can preliminarily use the patterns for legal compliance checking proposed by Francesconi and Governatori [51]. Yet, at the implementation level, their semantic distinction between *provisions* and *norms* could be fleshed out incorporating the pragmatic dimension that is needed to generate and validate legal ecosystems. At the microlevel, more variables should be considered to get them done in a sustainable way. The validation of the smart regulatory ecosystem should be data-driven. The accuracy of the validation is depending on the quality of the dataflow provided to feed the system.

*3.3. Third Step: A Compliance Causal Model*

To enable an *empirical approach* to legal sources, norms, and smart legal ecosystems, we could construct their causal chains (including computer and human behaviour). This is the third step. This involves building the causal-loop models [52] (i) learning and defining the degree of relationships and inter-dependence between various components of the regulatory ecosystem impacting the validity (positive or inhibitory effects), (ii) and modelling deeper (three-tier) levels of complexity of interactions in the legal governance model. Figure 4 draws the causal legal validation scheme from the components of the metamodel of legal governance and their relationships. These can be used in the regulatory simulation process.

The model could be tested, refined, and optimised in three different OPTIMAI 4.0 scenarios: (i) *quality checking* (multimodal sensor network allowing for smart and secure data collection on production lines); (ii) *augmented reality* (context-aware environment using AR glasses to optimise production chains); and (iii) *digital twins* (digital technology allowing the virtualisation of the production process). There are three use cases corresponding to three separate pilots.

The legal validation process, i.e. the generation of a sustainable ecosystem *which can be deemed legal*, can be performed (i) defining and fleshing out the conceptual scheme of Figure 1; (ii) implementing the meta-model dynamic process of Figure 2; (iii) and testing legal compliance through the causal model of Figure 3. The final outcome can be deemed the OPTIMAI regulatory model.

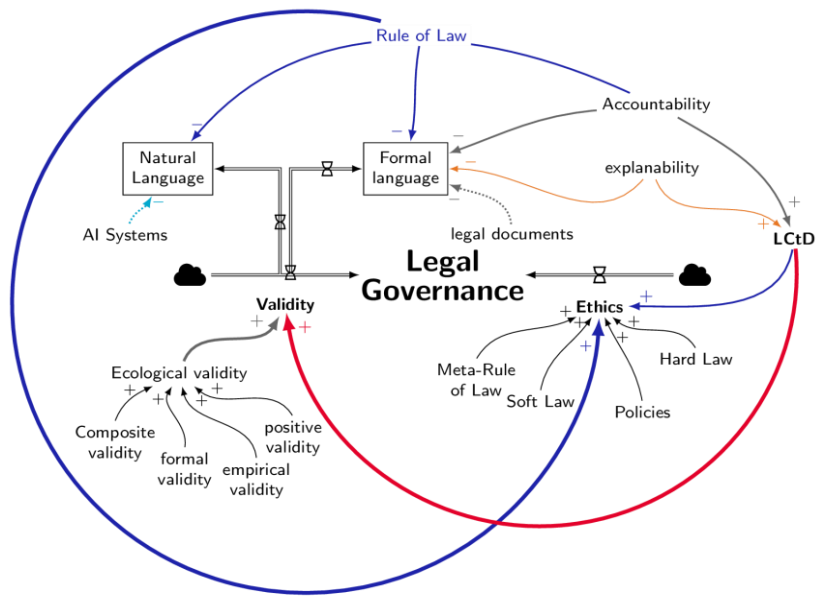

**Figure 4.** Legal Governance Meta-model: A Causal Legal Validation Scheme.

### 4. Conclusions and future work

Some years ago [53] we anticipated that law is facing significant new challenges, related to personalisation of web services, unregulated contexts and scenarios, emerging data markets, non-harmonised jurisdictions, safety, and collective security. We identified ten topics to be discussed. Among them, the relevance of ethics; the need to align social, legal, and technological knowledge; and the need to solve the algorithmic-semantic puzzle.

We have presented in this position paper a three-step methodology aimed at validating legal governance models from an empirical point of view: (i) a scheme for the rule and metarule of law; (ii) a metamodel for legal governance to be implemented by means of Compliance through Design (CtD); (iii) a compliance causal model to validate the generated smart legal ecosystem. *Legal validity* and *legal validation* processes are kept and treated in a separate analytical way, using a range of differentiated techniques.

This methodology can be developed and implemented in several distinct fields as well (such as security, health, and banking). In banking, for instance, some recurrent legal compliance problems such as (i) the identification and verification of clients' identity required by law (known as 'Know Your Customer/KYC' processes), (ii) the identification of transactions suspected of involving proceeds of crime, (iii) and the control of the export of goods that may have military use or civilian use (known as 'dual use goods'), could benefit from this tripartite approach.

We also identified some challenges. Among them: (i) norm and rule extraction (from natural language); (ii) the combination of documentary (written) and behavioural (oral) sources; (iii) the coexistence and coordination of a *dual-loop closed system* with its legal validation; (iv) the coexistence and coordination of interpretive (human) decisions with real-time execution of compliance processes. We can add the effort to build a usable concept of *ecological validity*. It is surprising that there is still no composite indicator for legal validity. It does not yet exist.

In the present position paper, to introduce and test our methodology for legal governance and compliance we have drawn from our work on OPTIMAI, a project of smart manufacturing bridging I4.0 and I5.0 and covering a wide range of production and distribution processes. OPTIMAI, a platform-driven information processing system, has built a *dual closed-loop system* on physical resources and supervisory control terminals, keeping humans-in-the loop. We are proposing a third normative loop to generate a smart legal ecosystem and a semi-automated legal validation process. This requires a closer attention to blockchain and smart contracts, the middleware system, and to the integration of data to feeding the regulatory system. The construction of the OPTIMAI regulatory model (ORM) will depend on these data analysis requirements, on the formal compliance language to substantiate ORM, and on the metrics that are also required to validate it. Ethical and legal sandboxes with all stakeholders are also required at this stage. In the immediate future, we can compare this approach with some results of the EU project MOSAIC. Qualitative reasoning and the possibility of using substructural modal logics to represent degrees of uncertainty can be collated with the degrees of rule compliance assumed in our approach.

*Acknowledgments*. The work presented in this paper has been partially funded by (i) the EU H2020 Program *Optimizing Manufacturing Processes through Artificial Intelligence and Virtualization* OPTIMAI (2021-2023), Grant agreement ID: 958264, (ii) the EUH2020 Project *Modalites in Substructural Logics: Theory, Methods and Applications* (MOSAIC) (2021-2026). Grant agreement ID: 101007627.